\documentclass[12pt]{article}
\usepackage{graphicx,fancyhdr}
\newcommand{\dfdx}[2]{\frac{\partial #1}{\partial #2}}
\begin{document}
\def\K{\mathord{\cal K}}
\def\la{\langle}
\def\ra{\rangle}
\def\ltsim{\mathop{\,<\kern-1.05em\lower1.ex\hbox{$\sim$}\,}}
\def\gtsim{\mathop{\,>\kern-1.05em\lower1.ex\hbox{$\sim$}\,}}

\begin{center}


\subsection*{ Many-body Green's
function theory for thin ferromagnetic films:
exact treatment of the single-ion anisotropy}
\vspace{2cm}

{\bf P. Fr\"obrich$^+$,  P.J. Kuntz, and M. Saber$^{++}$}
\vspace{1cm}

Hahn-Meitner-Institut Berlin, Glienicker Stra{\ss}e 100, D-14109 Berlin,
Germany,\\
$^+$also: Institut f\"ur Theoretische Physik, Freie Universit\"at Berlin\\
Arnimallee 14, D-14195 Berlin, Germany\\
$^{++}$ home address: Physics Department, Faculty of Sciences, University
Moulay Ismail, B.P. 4010 Meknes, Morocco
\end{center}
\vspace{3cm}

\noindent {\bf Abstract.}
A theory for the magnetization of  ferromagnetic films
is formulated within the framework of many-body Green's function theory
which considers all components of the magnetization.
The model Hamiltonian includes
a Heisenberg term, an external magnetic field, a
second- and fourth-order uniaxial single-ion anisotropy, and the magnetic
dipole-dipole coupling. The single-ion
anisotropy terms can be treated {\em exactly} by
introducing higher-order Green's functions and subsequently taking advantage of
relations between
products of spin operators which leads to an automatic closure of the
hierarchy of the equations of motion for the Green's functions with respect to
the anisotropy terms. This is an improvement on the method of our previous
work, which treated the corresponding terms only approximately
by decoupling them at the level of the lowest-order Green's functions.
RPA-like approximations are used to decouple the exchange interaction terms
in both the low-order and higher-order Green's functions.
As a first numerical example we apply the theory to a monolayer for spin
$S=1$ in order to demonstrate the superiority
of the present treatment of the anisotropy terms over the previous
approximate decouplings.

\noindent {\bf Keywords:}
Quantized spin model; Many-body Green's functions; thin ferromagnetic films.

\subsection*{1 Introduction}
There is increasing activity in experimental and theoretical investigations of
the properties of thin magnetic films and multi-layers. Of particular interest
is the magnetic reorientation transition which is measured as function of
temperature and film thickness; for recent papers, see
\cite{Sell01,Put01} and references therein.

The simplest theoretical approach to the treatment of thin
ferromagnetic films in the Heisenberg model is the mean field theory (MFT),
which can be applied
either by diagonalization of a single-particle Hamiltonian\cite{Mos94}
or by thermodynamic perturbation theory\cite{Jen96}. This
approximation, however, completely neglects collective excitations
(spin waves), which are known to be much more important
for the magnetic properties of 2D systems than for 3D bulk materials.
In order to take the influence of collective excitations into account,
one can turn to many-body Green's function theory (GFT), which allows
reliable calculations over the entire range of temperature of
interest: see, for example, Refs.~\cite{FJK00,FJKE01,Guo00}, where the
formalism includes the magnetic reorientation. The application of
Green's functions after a Holstein-Primakoff mapping to bosons, as
applied in Ref.~\cite{Eric91},
is valid only at low temperatures. Another method, which also can treat
the magnetic reorientation for all temperatures, is the application of a
Schwinger-Boson theory\cite{Tim00}. Classical Monte Carlo calculations
are also able to simulate the reorientation transition (see \cite{Mac98} and
references therein).  The temperature-dependent reorientation transition has
also been investigated with a Hubbard model\cite{Her98}.

In the present paper, we apply a Green's function theory to a Heisenberg
Hamiltonian plus anisotropy terms, a system previously treated at the level
of the lowest-order Green's functions\cite{FJK00,FJKE01,Guo00}. The
{\em approximate} treatment of the single-ion anisotropy in the previous work
is avoided here by extending the
formalism to higher-order Green's functions and applying relations for products
of spin operators, a procedure which leads to automatic closure
of the hierarchy of equations of motion with respect to those terms
stemming from the single-ion anisotropy.
The exchange terms occurring in the higher-order Green's functions
must, however, still be decoupled in an RPA-like fashion. This can be
considered as an extension of the work of Devlin\cite{Dev71}, who has
applied higher-order
Green's functions to the description of bulk magnetic materials in
one direction only. Our formulation applies to all spatial directions
of a multi-layer system.  We formulate the theory explicitly for
a monolayer for spin $S=1$ and give equations for an extension to
the multi-layer case. It is straightforward to see how the theory
could be applied to higher spins.

The paper is organized as follows: in Section~2, the previous
theory\cite{FJK00,FJKE01} for thin ferromagnetic
films is generalized by introducing higher-order Green's functions,
the model being explained in detail for a monolayer with $S=1$.
Section~3 gives the formal extension to the multi-layer case.
In Section~4, numerical results for the $S=1$ monolayer demonstrate
the superiority of the exact treatment of
the single-ion anisotropy term over the previously
applied\cite{FJK00,FJKE01}
Anderson-Callen\cite{AC64} decoupling. Section~5 contains a summary of the
results.

\subsection*{2 The Green's function formalism}

We investigate here a spin Hamiltonian, nearly the same
as in Ref.~\cite{FJKE01}, consisting
of an isotropic Heisenberg exchange interaction between nearest
neighbour lattice sites, $J_{kl}$, second- and fourth-order
single-ion lattice anisotropies with strengths $K_{2,k}$ and $K_{4,k}$
respectively, a magnetic dipole coupling with
strength $g_{kl}$, and an external magnetic field ${\bf{B}}=(B^x,B^y,B^z)$:
\begin{eqnarray}
{\cal H}&=&-\frac{1}{2}\sum_{<kl>}J_{kl}(S_k^-S_l^++S_k^zS_l^z)\nonumber\\
& &
-\sum_kK_{2,k}(S_k^z)^2-\sum_kK_{4,k}(S_k^z)^4\nonumber\\
& &
-\sum_k\Big(\frac{1}{2}B^-S_k^++\frac{1}{2}B^+S_k^-
+B^zS_k^z\Big) \nonumber\\
& & +\frac{1}{2}\sum_{kl}\frac{g_{kl}}{r_{kl}^5}\Big(r_{kl}^2(S_k^-S_l^+
+S_k^zS_l^z)-3({\bf{S}}_k\cdot{\bf{r}}_{kl})({\bf{S}}_l\cdot{\bf
r}_{kl})\Big)\ .
\label{1}
\end{eqnarray}
Here the notation
$S_i^\pm=S_i^x\pm iS_i^y$ and $B^\pm=B^x\pm iB^y$ is introduced, where
$k$ and $l$
are lattice site indices, and $\langle kl\rangle$ indicates summation
over nearest neighbours only. Here, we add to the Hamiltonian
in Ref.~\cite{FJKE01} a
fourth-order anisotropy term which we can treat {\em exactly}
but for which we previously had no decoupling procedure available.

Each layer is assumed to be ferromagnetically ordered: spins on each site
in the same layer are parallel, whereas spins belonging to different layers
need not be. Furthermore, the anisotropy strengths, coupling
constants and magnetic moments are considered to be layer-dependent, so
that inhomogeneous systems can be considered.

To allow as general a formulation as possible (with an eye to a future
study of the reorientation of the magnetization), we formulate the
equations of motion for the Green's functions for all spatial
directions:
\newpage
\begin{eqnarray}
G_{ij}^{+,\mp}(\omega)&=&\la\la S_i^+;S_j^{\mp}\ra\ra_\omega \nonumber \\
G_{ij}^{-,\mp}(\omega)&=&\la\la S_i^-;S_j^{\mp}\ra\ra_\omega  \\
G_{ij}^{z,\mp}(\omega)&=&\la\la S_i^z;S_j^{\mp}\ra\ra_\omega \ . \nonumber
\label{2}
\end{eqnarray}
Instead of decoupling the corresponding equations of motion at this stage,
as we did in our previous work\cite{FJK00,FJKE01}, we add
equations for the next higher-order Green's functions:

\begin{eqnarray}
\label{3}
G_{ij}^{z+,\mp}(\omega)&=&\la\la (2S_i^z-1)S_i^+;S_j^{\mp}\ra\ra_\omega
\nonumber
\\ G_{ij}^{-z,\mp}(\omega)&=&\la\la S_i^-(2S_i^z-1);S_j^{\mp}\ra\ra_\omega
\nonumber
\\ G_{ij}^{++,\mp}(\omega)&=&\la\la S_i^+S_i^+;S_j^{\mp}\ra\ra_\omega
\\ G_{ij}^{--,\mp}(\omega)&=&\la\la S_i^-S_i^-;S_j^{\mp}\ra\ra_\omega \nonumber
\\ G_{ij}^{zz,\mp}(\omega)&=&\la\la
(6S_i^zS_i^z-2S(S+1));S_j^{\mp}\ra\ra_\omega \ . \nonumber
\end{eqnarray}
The particular form for the operators used in the definition of the Green's
functions in Eqs.~(\ref{3}) is dictated by expressions coming
from the anisotropy terms. Terminating
the hierarchy of the equations of motion at this level of the Green's
functions results in an {\em exact} treatment of the
anisotropy terms for spin $S=1$, since the hierarchy for these terms
breaks off at this stage, as will be shown. The exchange interaction terms,
however, still have to be decoupled, which we do with RPA-like decouplings.

For the treatment of arbitrary spin $S$, it is necessary to use
$4S(S+1)$ Green's functions in order
to obtain an automatic break-off of the equations-of-motion hierarchy
coming from the anisotropy terms. These are functions of the structure
$G_{ij}^{\alpha,\mp}$ with $\alpha=(z)^n(+)^m$ and $\alpha=(-)^m(z)^n$, where,
for a particular spin $S$, all combinations of $m$ and $n$ satisfying
$(n+m)=2S$ have to be taken into account. There occur no Green's functions
having mixed $+$ and $-$ indices, because these can be
reduced by the relation $S^\mp S^\pm=S(S+1)\mp S^z-(S^z)^2$.

The equations of motion which determine the Green's functions are
\begin{equation}
\omega G_{ij}^{\alpha,\mp}=A_{i,j}^{\alpha,\mp}+\la\la [O_i^\alpha,{\cal
H}]_-;S_j^\mp\ra\ra,
\label{4}
\end{equation}
with the inhomogeneities
\begin{equation}
A_{ij}^{\alpha,\mp}=\la [O_i^\alpha,S_j^\mp]_-\ra,
\label{5}
\end{equation}
where $O_i^\alpha$ are the operators occuring in the definition of the Green's
functions, and $\la ... \ra=Tr(...e^{-\beta {\cal H}})$.
\newpage
In the following, we treat a monolayer with $S=1$ explicitly. In this case,
a system of 8 equations of motion is necessary:
\begin{eqnarray}
\label{6}
\omega G_{ij}^{+,\mp}&=&A_{ij}^{+,\mp}-\sum_k J_{ik}\la\la
S_i^zS_k^+-S_k^zS_i^+;S_j^\mp\ra\ra
+B^zG_{ij}^{+,\mp}-B^+G_{ij}^{z,\mp}\nonumber\\
& &+(K_{2,i}+K_{4,i})G_{ij}^{z+,\mp}-2K_{i,4}(G_{ij}^{(z)^2+,\mp}-
G_{ij}^{(z)^3+,\mp}),\nonumber\\
\omega G_{ij}^{-,\mp}&=&A_{ij}^{-,\mp}+\sum_k J_{ik}\la\la
S_i^zS_k^--S_k^zS_i^-;S_j^\mp\ra\ra
-B^zG_{ij}^{-,\mp}+B^-G_{ij}^{z,\mp}\nonumber\\
& &-(K_{2,i}+K_{4,i})G_{ij}^{-z,\mp}+2K_{i,4}(G_{ij}^{-(z)^2,\mp}-
G_{ij}^{-(z)^3,\mp}),\nonumber\\
\omega G_{ij}^{z,\mp}&=&A_{ij}^{z,\mp}+\frac{1}{2}\sum_k J_{ik}\la\la
S_i^-S_k^+-S_k^-S_i^+;S_j^\mp\ra\ra
-\frac{1}{2}B^-G_{ij}^{+,\mp}+\frac{1}{2}B^+G_{ij}^{-,\mp},\nonumber\\
\omega G_{ij}^{z+,\mp}&=&A_{ij}^{z+,\mp}-\frac{1}{2}\sum_k J_{ik}\Big(\la\la
(6S_i^zS_i^z-2S(S+1))S_k^+;S_j^\mp\ra\ra\nonumber\\
& &+2\la\la S_k^-S_i^+S_i^+;S_j^\mp\ra\ra
-2\la\la S_k^z(S_i^zS_i^++S_i^+S_i^z);S_j^\mp\ra\ra\Big)\nonumber\\
& &-B^-G_{ij}^{++,\mp}-\frac{1}{2}B^+G_{ij}^{zz,\mp}
+B^zG_{ij}^{z+,\mp}\nonumber\\
& &-(K_{2,i}+K_{4,i})G_{ij}^{z+,\mp}+(2K_{2,i}+4K_{4,i})G_{ij}^{(z)^2+,\mp}
-6K_{i,4}G_{ij}^{(z)^3+,\mp}+4K_{4,i}G_{ij}^{(z)^4+,\mp},\nonumber\\
\omega G_{ij}^{-z,\mp}&=&A_{ij}^{-z,\mp}+\frac{1}{2}\sum_k J_{ik}\Big(\la\la
S_k^-(6S_i^zS_i^z-2S(S+1));S_j^\mp\ra\ra\nonumber\\
& &+2\la\la S_i^-S_i^-S_k^+;S_j^\mp\ra\ra
-2\la\la S_k^z(S_i^zS_i^-+S_i^-S_i^z);S_j^\mp\ra\ra\Big)\nonumber\\
& &+\frac{1}{2}B^-G_{ij}^{zz,\mp}+B^+G_{ij}^{--,\mp}
-B^zG_{ij}^{-z,\mp}\nonumber\\
& &+(K_{2,i}+K_{4,i})G_{ij}^{-z,\mp}-(2K_{2,i}+4K_{4,i})G_{ij}^{-(z)^2,\mp}
+6K_{i,4}G_{ij}^{-(z)^3,\mp}-4K_{4,i}G_{ij}^{-(z)^4,\mp},\nonumber\\
\omega G_{ij}^{++,\mp}&=&A_{ij}^{++,\mp}-\sum_k J_{ik}\Big(\la\la
(S_i^zS_i^++S_i^+S_i^z)S_k^+;S_j^\mp\ra\ra-2\la\la
S_i^+S_i^+S_k^z;S_j^\mp\ra\ra\Big)\nonumber\\
& &-B^+G_{ij}^{z+,\mp}+2B^zG_{ij}^{++,\mp}
-4K_{2,i}(G_{ij}^{++,\mp}-G_{ij}^{z++,\mp})\nonumber\\
& &-K_{i,4}(8G_{ij}^{(z)^3++,\mp}-
24G_{ij}^{(z)^2++,\mp}+32G_{ij}^{z++,\mp}-16G_{ij}^{++,\mp}),\nonumber\\
\omega G_{ij}^{--,\mp}&=&A_{ij}^{--,\mp}+\sum_k J_{ik}\Big(\la\la
S_k^-(S_i^zS_i^-+S_i^-S_i^z);S_j^\mp\ra\ra-2\la\la
S_i^-S_i^-S_k^z;S_j^\mp\ra\ra\Big)\nonumber\\
& &+B^-G_{ij}^{-z,\mp}-2B^zG_{ij}^{--,\mp}
+4K_{2,i}(G_{ij}^{--,\mp}-G_{ij}^{--z,\mp})\nonumber\\
& &-K_{i,4}(8G_{ij}^{--(z)^3,\mp}-
24G_{ij}^{--(z)^2,\mp}+32G_{ij}^{--z,\mp}-16G_{ij}^{--,\mp}),\nonumber\\
\omega G_{ij}^{zz,\mp}&=&A_{ij}^{zz,\mp}+\sum_k J_{ik}\Big(3\la\la
(S_i^zS_i^-+S_i^-S_i^z)S_k^+;S_j^\mp\ra\ra -3\la\la
S_k^-(S_i^zS_i^++S_i^+S_i^z);S_j^\mp\ra\ra\Big)\nonumber\\
& &-3B^-G_{ij}^{z+,\mp}+3B^+G_{ij}^{-z,\mp}.
\end{eqnarray}
These equations are exact. The important point now is that
the anisotropy terms in these equations can be simplified by using
formulae which reduce  products of spin operators by one order. Such relations
were derived in Ref.~\cite{JA2000}:
\begin{eqnarray}
\label{7}
(S^-)^m(S^z)^{2S+1-m}&=&(S^-)^m\sum_{i=0}^{2S-m}\delta_i^{(S,m)}(S^z)^i,
\nonumber\\
(S^z)^{2S+1-m}(S^+)^m&=&\sum_{i=0}^{2S-m}\delta_i^{(S,m)}(S^z)^i(S^+)^m.
\end{eqnarray}
The coefficients $\delta_i^{(S,m)}$ are tabulated in Ref.~\cite{JA2000}
for general spin.
For spin $S=1$, only the coefficients with $m=0,1,2$ occur:
$\delta_0^{(1,0)}=\delta_2^{(1,0)}=0; \delta_1^{(1,0)}=1, \delta_0^{(1,1)}=0,
\delta_1^{(1,1)}=1, \delta_0^{(1,2)}=1$.

Application of these relations, effects the reduction of
the relevant Green's functions coming from the anisotropy terms in
equations (\ref{6}):
\begin{eqnarray}
\label{8}
G_{ij}^{(z)^4+,\mp}&=&G_{ij}^{(z)^3+,\mp}=G_{ij}^{(z)^2+,\mp}
=\frac{1}{2}(G_{ij}^{z+,\mp} + G_{ij}^{+,\mp}),\nonumber\\
G_{ij}^{-(z)^4,\mp}&=&G_{ij}^{-(z)^3,\mp}=G_{ij}^{-(z)^2,\mp}
=\frac{1}{2}(G_{ij}^{-z,\mp} + G_{ij}^{-,\mp}),\nonumber\\
G_{ij}^{(z)^2++,\mp}&=&G_{ij}^{z++,\mp}=G_{ij}^{++,\mp},\\
G_{ij}^{--(z)^2,\mp}&=&G_{ij}^{--z,\mp}=G_{ij}^{--,\mp}.\nonumber
\end{eqnarray}
The higher Green's functions coming from the
anisotropy terms have thus been expressed in terms of the lower-order functions
already present in the hierarchy; i.e. with respect to the anisotropy terms,
a closed system of equations of motion results, so that no decoupling of
these terms is necessary.  In other words, the anisotropy is treated
{\em exactly}. For higher spins, $S>1$, one can proceed
analogously. For this, one needs even higher-order Green's functions but again,
applying equations (\ref{7}) reduces the relevant Green's functions
by one order, which in turn leads to a closed system of equations
obviating the decoupling of terms coming from the anisotropies.

No such procedure is available for the exchange interaction terms, however,
so that these still have to be decoupled. For spin $S=1$,
we use RPA-like approximations to effect the decoupling:
\begin{eqnarray}
\la\la S_i^\alpha S_k^\beta;S_j^\mp\ra\ra&\simeq&
\la S_i^\alpha\ra G_{kj}^{\beta,\mp}+\la S_k^\beta\ra
G_{ij}^{\alpha,\mp}\nonumber \\
\la\la S_k^\alpha S_i^\beta S_i^\gamma;S_j^\mp\ra\ra&\simeq&
\la S_k^\alpha\ra G_{ij}^{\beta\gamma,\mp}+\la S_i^\beta S_i^\gamma \ra
G_{kj}^{\alpha,\mp}.
\label{9}
\end{eqnarray}
Note that we have constructed the decoupling so as not to
break correlations having equal indices, since the corresponding
operators form the algebra characterizing the group for a
spin $S=1$ system.
 For spin $S=1$, this decoupling model leads to 8
diagonal correlations for each layer $i$:
\begin{center}
$\la S_i^+\ra, \la S_i^-\ra, \la S_i^z\ra,$
$ \la S_i^+S_i^+\ra, \la S_i^-S_i^-\ra, \la S_i^zS_i^z\ra,
\la S_i^zS_i^+\ra, \la S_i^-S_i^z\ra, \la S_i^zS_i^z\ra.$
\end{center}
These have to be determined by $8\times N$ equations, where $N$ is the number
of layers. We have not attempted to go beyond the RPA-approximation because a
previous comparison of Green's function theory with `exact' quantum
Monte Carlo calculations for a Heisenberg hamiltonian for a monolayer with
$S=1/2$ in a magnetic field showed RPA to be a remarkably good
approximation\cite{EFJK99}.

We now apply the above reduction, Eqs.~(\ref{8}), and the decoupling of the
exchange interaction terms, Eqs.~(\ref{9}), to the {\em monolayer}
with  spin $S=1$.
Then, after performing a two-dimensional Fourier transformation,
one obtains a set of equations of motion, which, in compact
matrix notation (dropping the layer index), is as follows:
\begin{equation}
(\omega {\bf 1}-{\bf \Gamma}){\bf G}^{\mp}={\bf A}^{\mp},
\label{10}
\end{equation}
where ${\bf G}^\mp$ and ${\bf A}^\mp$ are 8-dimensional vectors with
components
$G^{\alpha,\mp}$ and $A^{\alpha,\mp}$ where $\alpha=+,-,z,z+,-z,++,--,zz$, and
${\bf 1}$ is the unit matrix. The $8\times 8$ {\it non-symmetric}
matrix ${\bf \Gamma}$ is given by
\begin{equation}
\scriptsize{
{\bf \Gamma} = \left( \begin{array}
{@{\hspace*{3mm}}c@{\hspace*{5mm}}c@{\hspace*{5mm}}c@{\hspace*{3mm}}
@{\hspace*{3mm}}c@{\hspace*{3mm}}c@{\hspace*{5mm}}c@{\hspace*{3mm}}
@{\hspace*{3mm}}c@{\hspace*{3mm}}c@{\hspace*{3mm}}}
\;\;\;\; H^z_k & 0 & -H^+_k & \tilde{K_2} & 0 & 0 & 0 & 0 \\
 0 & -H^z_k& \;\;\;H^-_k & 0 & -\tilde{K_2} & 0 & 0 & 0 \\
-\frac{1}{2}H^-_k & \;\frac{1}{2}H^+_k & 0 & 0 & 0 & 0 & 0 & 0 \\
\tilde{K_2}-\frac{J_k}{2}\la 6S^zS^z-4\ra & -\la S^+S^+\ra J_k &
\la(2S^z-1)S^+\ra J_k&H^z&0&-H^-&0&-\frac{1}{2}H^+\\
\la S^-S^-\ra J_k & -\tilde{K_2}+\frac{J_k}{2}\la 6S^zS^z-4\ra &-\la
S^-(2S^z-1)\ra J_k & 0 & -H^z & 0 & H^+ & \frac{1}{2}H^- \\
 -\la(2S^z-1)S^+\ra J_k & 0 & 2\la S^+S^+\ra J_k & -H^+ & 0 & 2H^z & 0 & 0 \\
0 & \la S^-(2S^z-1)\ra J_k & -2\la S^-S^-\ra J_k & 0 & H^- & 0 & -2H^z & 0 \\
3\la S^-(2S^z-1)\ra J_k & -3\la(2S^z-1)S^+\ra J_k & 0 & -3H^- & 3H^+ & 0 & 0& 0
 \end{array} \right) \;,}
\label{11}
\end{equation}
with the abbreviations
\begin{eqnarray}
H^\alpha_k&=&B^\alpha+\la S^\alpha\ra J(q-\gamma_{\bf k})\,,
\qquad \alpha=+,-,z \nonumber\\
H^\alpha&=&B^\alpha+\la S^\alpha\ra Jq, \qquad \alpha=+,-,z
\nonumber\\ J_k&=&J\gamma_{\bf k},
\label{12} \\
\tilde{K_2}&=&K_2+K_4.\nonumber
\end{eqnarray}
For a square lattice with a lattice constant taken to be unity,
$\gamma_{\bf k}=2(\cos k_x+\cos k_y)$, and $q=4$, the number of
nearest neighbours.
For spin $S=1$ and $S=3/2$, the $K_4$ term in the Hamiltonian leads only to a
renormalization of the second-order anisotropy coefficient:
$\tilde{K_2}(S=1)=K_2+K_4$, and $\tilde{K_2}(S=3/2)=K_2+\frac{5}{2}K_4$
respectively. Only in the case of higher spins,
$S\geq 2$, are there non-trivial corrections due to the fourth-
order anisotropy coefficient.

If the theory is formulated only in terms of  $G^-$, there is no equation
for determining the $\la S^+S^+\ra$ occuring in the  $\bf \Gamma-$matrix.
It is for this reason that we introduced $G^+$ in Eq.~(\ref{2}),
for which the $\bf \Gamma-$matrix turns out to be
the same, so that, in general, one can take a linear
combination of $G^+$ and $G^-$ and their corresponding inhomogeneities:
\begin{eqnarray}
{\bf G}&=&(1-a){\bf G}^{-}+a{\bf G}^{+},\nonumber\\
{\bf A}&=&(1-a){\bf A}^{-}+a{\bf A}^{+}.\nonumber\\
\label{13}
\end{eqnarray}
Hence, the equations of motion are
\begin{equation}
(\omega {\bf 1}-{\bf \Gamma}){\bf G}={\bf A},
\label{14}
\end{equation}
from which the desired correlations
${\bf C}=(1-a){\bf C}^{-}+a{\bf C}^{+}$
can be determined. The  parameter $a$ is arbitrary ($0<a<1$).
The correlation vector for spin $S=1$ in terms of the 8 correlations
mentioned above is
\begin{eqnarray}
{\bf C} & = &
 \left( \begin{array}{c}
\la((1-a)S^-+aS^+)S^+\ra \\ \la((1-a)S^-+aS^+)S^-\ra \\ \la((1-a)S^-+aS^+)S^z\ra
\\\la((1-a)S^-+aS^+)(2S^zS^+-S^+)\ra \\
\la((1-a)S^-+aS^+)(2S^-S^z-S^-)\ra \\ \la((1-a)S^-+aS^+)S^+S^+\ra \\
\la((1-a)S^-+aS^+)S^-S^-\ra \\ \la((1-a)S^-+aS^+)(6S^zS^z-4)\ra \nonumber
\end{array} \right) \\
        & = &
\left( \begin{array}{c}
(1-a)(2-\la S^z\ra-\la S^zS^z\ra)+a\la S^+S^+\ra \\
(1-a)\la S^-S^-\ra+a(2+\la S^z\ra-\la S^zS^z\ra) \\
(1-a)\la S^-S^z\ra+a(\la S^zS^+\ra-\la S^+\ra) \\
(1-a)(2+\la S^z\ra-3\la S^zS^z\ra)-a\la S^+S^+\ra\\
(1-a)\la S^-S^-\ra+a(\la S^z\ra+3\la S^zS^z\ra-2) \\
(1-a)(2\la S^+\ra-2\la S^zS^+\ra)+a\la S^+S^+S^+\ra\\
(1-a)\la S^-S^-S^-\ra+a2\la S^-S^z\ra\\
(1-a)(6\la S^-S^z\ra -4\la S^-\ra)+a(2\la S^+\ra-6\la S^zS^+\ra)
\end{array}
\right),
\label{15}
\end{eqnarray}
where one can introduce the identity (for spin $S=1$):
$\la S^+S^+S^+\ra=\la S^-S^-S^-\ra=0$.

The inhomogeneity vectors for spin $S=1$ are given by
\begin{equation}
 \left( \begin{array}{c}
A^{+,-} \\ A^{-,-}\\A^{z,-}\\A^{z+,-}\\A^{-z,-}\\A^{++,-}\\A^{--,-}\\A^{zz,-}
\end{array} \right)=
\left( \begin{array}{c}
2\la S^z\ra \\ 0 \\-\la S^-\ra \\ 6\la S^zS^z\ra-4 \\-2\la S^-S^-\ra \\4\la
S^zS^+\ra-2\la S^+\ra\\ 0\\ 6\la S^-\ra -12\la S^-S^z\ra \end{array} \right),
\hspace{0.5cm}
 \left( \begin{array}{c}
A^{+,+} \\ A^{-,+}\\A^{z,+}\\A^{z+,+}\\A^{-z,+}\\A^{++,+}\\A^{--,+}\\A^{zz,+}
\end{array} \right)=
\left( \begin{array}{c}
0\\ -2\la S^z\ra \\ \la S^+\ra \\ 2\la S^+S^+\ra \\4-6\la S^zS^z\ra \\ 0 \\
2\la
S^-\ra-4\la S^-S^z\ra\\ 12\la S^zS^+\ra -6\la S^+\ra \end{array} \right).
\label{16}
\end{equation}
The correlations ${\bf C}$ are related to the Green's functions via the
spectral theorem. In order to determine these, we apply the eigenvector
method already used in Ref.~\cite{FJKE01} and explained there in detail.
This method is
quite general and not restricted to the $8\times 8$ problem above; it also
makes the extension of the theory to multi-layer systems tractable.

The essential steps in deriving the coupled integral equations for determining
the correlations ${\bf C}$ are now outlined. One starts by diagonalizing
the {\em non-symmetric} matrix ${\bf \Gamma}$ of equation
(\ref{14})
\begin{equation}
{\bf L\Gamma R}={\bf \Omega},
\label{17}
\end{equation}
where ${\bf R}$ is a matrix whose columns are the right eigenvectors of
${\bf \Gamma}$ and its inverse ${\bf L=R^{-1}}$ contains the left
eigenvectors as rows, where ${\bf RL=LR=1}$. Multiplying Eq.~(\ref{14})
from the left by ${\bf L}$ and inserting ${\bf RL=1}$ yields
\begin{equation}
(\omega {\bf 1}-{\bf \Omega}){\cal G}={\cal A},
\label{18}
\end{equation}
where we introduce ${\cal G}={\bf LG}$ and ${\cal A}={\bf LA}$.
Here ${\cal G}$ is a new vector of Green's functions with the property
that each component ${\cal G}^\tau$ has but a single pole
\begin{equation}
{\cal G}^\tau =\frac{\cal A}{\omega-\omega_\tau}.
\label{19}
\end{equation}
This allows the application of the spectral theorem\cite{Elk01}
to each component separately, with ${\cal C}={\bf LC}$:

\begin{equation}
{\cal C}^\tau =\frac{{\cal A}_{\eta=-1}}{e^{\beta\omega_\tau}-1}+{\cal D}^\tau.
\label{20}
\end{equation}
${\bf \cal D}={\bf LD}$ is the correction to the
spectral theorem needed in case there are vanishing eigenvalues. The
corresponding components ${\cal D}^\tau$ are
obtained from the anticommutator Green's function
${\cal G}_{\eta=+1}$:
\begin{equation}
{\cal D}^\tau =\frac{1}{2}\lim_{\omega\rightarrow 0}\omega {\cal G}_{\eta=+1}=
\frac{1}{2}\lim_{\omega\rightarrow 0}\frac{\omega{\cal
A}_{\eta=+1}}{\omega-\omega_\tau} =\frac{1}{2}\delta_{\omega_\tau 0}{\cal
A}^0_{\eta=+1}\ ,
\label{21}
\end{equation}
i.e. ${\cal D}^\tau$ is non-zero only for eigenvalues $\omega_\tau=0$.
Denoting these by ${\cal D}^0$ and the corresponding left eigenvectors
by ${\bf L^0}$, one obtains from the Eq.~(\ref{21})
\begin{equation}
{\cal D}^0=\frac{1}{2}{\cal A}^0_{\eta=+1}=\frac{1}{2}{\bf L^0A}_{\eta=+1}=
\frac{1}{2}{\bf L^0}({\bf A}_{\eta=-1}+2{\bf C})={\cal C}^0.
\label{22}
\end{equation}
Here, we have exploited the fact that the
commutator Green's function is regular at the origin
(called the regularity condition in \cite{FJKE01}):
\begin{equation}
{\bf L}^0{\bf A}_{\eta=-1}=\sum_{\alpha}L_{\alpha}^0A^{\alpha}_{\eta=-1}=0.
\label{23}
\end{equation}
The desired correlation vector ${\bf C}$ is now obtained by multiplying the
correlation vector ${\cal C}$, Eq.~(\ref{20}), from the left by ${\bf R}$:

\begin{equation}
{\bf C}={\bf R\;}{\bf \cal E\;}{\bf L\;}{\bf A}+{\bf R}^0{\bf L}^0{\bf C}.
\label{24}
\end{equation}
Here, the two terms on the right-hand side belong to the non-zero
and zero eigenvalues of the $\bf \Gamma-$matrix, respectively.
${\bf R}$ is the matrix whose columns are the right eigenvectors
of the ${\bf \Gamma}$-matrix with
eigenvalues $\omega_\tau\neq 0$  and
${\bf L}$ is the corresponding matrix whose rows are the left
eigenvectors with eigenvalues $\omega_\tau\neq 0$. ${\bf \cal E}$ is
a diagonal matrix whose elements are the functions
$\frac{1}{\exp{(\beta\omega_\tau)}-1}$. The matrices ${\bf R}^0$
and ${\bf L}^0$ consist of  the right (column) and left (row) eigenvectors
corresponding to eigenvalues $\omega_\tau=0$.
This constitutes a system of integral equations which has to be solved
self-consistently.

Note that the right-hand side of Eq.~(\ref{24}) contains
a Fourier transformation,
which can be made manifest by writing out the equations for each component
of $\bf C$ explicitly:
\begin{equation}
{C}_i=\frac{1}{\pi^2}\int_0^\pi
dk_x\int_0^\pi dk_y \sum_{l=1}^N\Big(\sum_{j=1}^n\sum_{k=1}^n
R_{ij}{\cal E}_{jk}\delta_{jk}L_{kl}{A}_l+\sum_{j=1}^m
R_{ij}^0L_{jl}^0{C}_l\Big).
\label{25}
\end{equation}
Here we have $i=1,..,N$ correlations ${C}_i$ corresponding to the
N-dimensional ${\bf \Gamma}$-
matrix with $n$ non-zero and $m$ zero eigenvalues ($n+m=N$). The momentum
integral goes over the first Brillouin zone.
For the case of a monolayer with spin $S=1$, the total number
of eigenvalues is $N=8$,
and one can show, by writing down the characteristic equation of the
$\bf \Gamma-$matrix, that 2 eigenvalues are exactly zero; i.e. $n=6, m=2$.

In general this matrix equation can be ill-defined, for, without
loss of generality, one can choose the field component $B^y$ to be zero,
in which case the correlations $\la (S^z)^m(S^+)^n\ra$ are the same as
$\la (S^-)^n(S^z)^m\ra$. This leads to a system of overdetermined equations.
These equations are solved by means of
a singular value decomposition\cite{Press89}, which is now illustrated
for spin $S=1$. In this case, we have
$\la S^+\ra=\la S^-\ra, \la S^+S^+\ra=\la S^-S^-\ra$, and
$\la S^zS^+\ra=\la S^-S^z\ra$; i.e. there are only 5 independent variables
defining the 8 correlations $\bf C$. We denote these
variables by the vector ${\bf v}$
\begin{equation}
{\bf v}=
 \left( \begin{array}{c}
\la S^-\ra \\ \la S^z\ra \\ \la S^-S^-\ra\\ \la S^-S^z\ra \\
\la S^zS^z\ra
\end{array} \right)
\label{100}
\end{equation}
Then, the correlations $\bf C$ can be expressed as
\begin{equation}
{\bf C}={\bf u_c^0}+{\bf u_cv}
\label{101}
\end{equation}
with
\begin{equation}
{\bf u_c^0}=
 \left( \begin{array}{c}
2-2a \\ 2a \\ 0 \\ 2-2a \\ -2a \\ 0 \\ 0\\ 0
\end{array} \right)
\label{102}
\end{equation}
and

\begin{equation}
{\bf u_c} = \left( \begin{array}
{@{\hspace*{3mm}}c@{\hspace*{5mm}}c@{\hspace*{5mm}}c@{\hspace*{3mm}}
@{\hspace*{3mm}}c@{\hspace*{3mm}}c@{\hspace*{5mm}}c@{\hspace*{3mm}}
@{\hspace*{3mm}}c@{\hspace*{3mm}}c@{\hspace*{3mm}}}
0 & a-1 & a & 0 & a-1 \\
0 & a & 1-a & 0 & -a \\
-a & 0 & 0 & 1 & 0 \\
0 & 1-a & -a & 0 & 3a-3 \\
0 & a & 1-a & 0 & 3a \\
2-2a & 0 & 0 & 2a-2 & 0 \\
0 & 0 & 0 & 2a & 0 \\
6a-4 & 0 & 0 & 6-12a & 0 \\
 \end{array} \right) \;.
\label{103}
\end{equation}
Now we write the $8\times5$ matrix ${\bf u_c}$ in terms of its
singular value decomposition:
\begin{equation}
{\bf u_c = U w \tilde{V}},
\label{104}
\end{equation}
where ${\bf w}$ is a $5\times 5$ diagonal matrix whose elements are
referred to as the singular values. These are in general zero or positive
but in our case they are all $>0$ for $0<a<1$.
${\bf U}$ is an orthogonal $8\times 5$ matrix
and ${\bf V}$ is a $5\times 5$ orthogonal matrix. From Eqs.~(\ref{24})
and~(\ref{101}) it follows that
\begin{equation}
{\bf u_cv=R\;{\bf \cal E\;} L\;A\; + R^0L^0(u_cv+u_c^0)-u_c^0}.
\label{105}
\end{equation}
 To get $\bf v$ from this equation, we need only multiply through by
${\bf u_c^{-1}=Vw^{-1}\tilde{U}}$, which yields
the system of coupled integral equations
\begin{equation}
{\bf v= u_c^{-1}\Big( R\;{\bf \cal E}\;L\;A +
R^0L^0(u_cv+u_c^0)-u_c^0}\Big), \label{106}
\end{equation}
or more explicitly with $i=1,...,5$
\begin{eqnarray}
v_i&=&\sum_{k=1}^8(u_c^{-1})_{ik}\frac{1}{\pi^2}\int_0^\pi dk_x\int_0^\pi dk_y
\sum_{j=1}^8\Big\{ \sum_{l=1}^6R_{kl}{\cal E}_{ll}L_{lj}A_j\nonumber\\
& &\ \ \  +\sum_{l=1}^2R_{kl}^0L_{lj}(\sum_{p=1}^5(u_c)_{jp}v_p+(u_c^0)_j
\Big\} -\sum_{k=1}^8(u_c^{-1})_{ik}(u_c^0)_k.
\label{107}
\end{eqnarray}
This set of equations is not overdetermined (5 equations for 5
unknowns in the
present example ) and is solved by the curve-following method described in
Appendix A.

\subsection*{3 The multilayer case}
For a ferromagnetic film with $N$ layers and spin $S=1$ one obtains   $8N$
equations of motion for the $8N$-dimensional Green's function vector ${\bf G}$
\begin{equation}
(\omega {\bf 1}-{\bf \Gamma}){\bf G},
\label{119}
\end{equation}
where ${\bf 1}$ is the $8N\times 8N$ unit matrix, and the Green's function and
inhomogeneity vectors consist of $N$ 8-dimensional subvectors which are
characterized by layer indices $i$ and $j$
\begin{eqnarray}
G_{ij}^\alpha(k,\omega)&=&(1-a)G_{ij}^{\alpha,+}+aG_{ij}^{\alpha,-},\nonumber\\
A_{ij}^\alpha(k,\omega)&=&(1-a)A_{ij}^{\alpha,+}+aA_{ij}^{\alpha,-}.
\label{120}
\end{eqnarray}
The equations of motion are then expressed in terms of these layer vectors, and
$8\times 8 $ submatrices ${\bf \Gamma}_{ij}$ of the $8N\times 8N$
matrix ${\bf\Gamma}$
\begin{equation}
\left[ \omega {\bf 1}-\left( \begin{array}{cccc}
{\bf\Gamma}_{11} & {\bf\Gamma}_{12} & \ldots & {\bf\Gamma}_{1N} \\
{\bf\Gamma}_{21} & {\bf\Gamma}_{22} & \ldots & {\bf\Gamma}_{2N} \\
\ldots & \ldots & \ldots & \ldots \\
{\bf\Gamma}_{N1} & {\bf\Gamma}_{N2} & \ldots & {\bf\Gamma}_{NN}
\end{array}\right)\right]\left[ \begin{array}{c}
{\bf G}_{1j} \\ {\bf G}_{2j} \\ \ldots \\ {\bf G}_{Nj} \end{array}
\right]=\left[ \begin{array}{c}
{\bf A}_{1j}\delta_{1j} \\ {\bf A}_{2j}\delta_{2j} \\ \ldots \\
{\bf A}_{Nj}\delta_{Nj} \end{array}
\right] \;, \hspace{0.5cm} j=1,...,N\;.
\label{121}
\end{equation}
In the multilayer case,
the  ${\bf \Gamma}$ matrix reduces to a band matrix with zeros in the
${\bf \Gamma}_{ij}$ sub-matrices, when $j>i+1$ and $j<i-1$.
The  diagonal sub-matrices ${\bf \Gamma}_{ii}$ are of size $8\times 8$
and have the same structure
as the matrix which characterizes the monolayer, see Eq.~(\ref{11}).
The matrix elements of ${\bf\Gamma}_{ii}$ contain terms depending
 on the layer index
$i$ and additional terms due to the exchange interaction between the atomic
layers.
\begin{eqnarray}
H^\alpha_{k,i}&=&B^\alpha_i+\la S_i^\alpha\ra J_{ii}(q-\gamma_{\bf k})
+J_{i,i+1}\la S_{i+1}^{\alpha}\ra+J_{i,i-1}\la S_{i-1}^{\alpha}\ra \,,
\qquad \alpha=+,-,z \nonumber \\
H^\alpha_{i}&=&B^\alpha_i+\la S_i^\alpha\ra J_{ii}q
+J_{i,i+1}\la S_{i+1}^{\alpha}\ra+J_{i,i-1}\la S_{i-1}^{\alpha}\ra \,,
\qquad \alpha=+,-,z \nonumber \\
\label{122}
\end{eqnarray}
The dipole coupling is treated in the mean field limit, which was
shown to be a good approximation for coupling strengths much weaker
than the exchange coupling\cite{FJKE01}. In this case, the dipole terms make
additive contributions to the magnetic field components
$B^\alpha_i$,
\begin{eqnarray}
B^{\pm}_i & \to & B^{\pm}+\sum_{j=1}^N\;g_{ij}\;\la S_j^{\pm}\ra\;
T^{|i-j|} \;, \nonumber\\
B^z_i & \to & B^z-2\sum_{j=1}^N\;g_{ij}\;\la S_j^z\ra\;
T^{|i-j|} \;,
\label{123}
\end{eqnarray}
where the lattice sums for a two-dimensional square lattice are given by

\begin{equation}
T^n = \sum_{lm}\frac{l^2-n^2}{(l^2+m^2+n^2)^{5/2}} \;,
\label{124}
\end{equation}
where $n=|i-j|$.
The indices ($lm$)  run over all sites of the square
$j$th layer, excluding the terms with $l^2+m^2+n^2=0$.
For the monolayer ($N=1$), one has $i=j$, and obtains
in particular $T^0\simeq 4.5165$.
As can be seen from Eqs.~(\ref{123}),
the dipole coupling reduces the effect of the external field component
in $z$-direction and enhances the effect of the transverse field
components $B^\pm$.

The $8\times 8$
non-diagonal sub-matrices ${\bf \Gamma}_{ij}$ for $j= i\pm 1$ are of the form
\begin{equation}
\scriptsize{
{\bf \Gamma}_{ij} = J_{ij}\left( \begin{array}
{@{\hspace*{3mm}}c@{\hspace*{3mm}}c@{\hspace*{3mm}}c@{\hspace*{3mm}}
@{\hspace*{3mm}}c@{\hspace*{3mm}}c@{\hspace*{3mm}}c@{\hspace*{3mm}}
@{\hspace*{1mm}}c@{\hspace*{3mm}}c@{\hspace*{3mm}}}
 -\la S_i^z\ra & 0 & \la S_i^+\ra & 0 & 0 & 0 & 0 & 0 \\
 0 & \la S_i^z\ra & -\la S_i^-\ra & 0 & 0 & 0 & 0 & 0 \\
\frac{1}{2}\la S_i^-\ra & -\frac{1}{2}\la S_i^+\ra & 0 & 0 & 0 & 0 & 0 & 0 \\
-\frac{1}{2}\la 6S^z_iS^z_i-4\ra & -\la S^+_iS^+_i\ra & \la(2S^z_i-1)S^+_i\ra
& 0 & 0 & 0 & 0 & 0 \\
\la S^-_iS^-_i\ra & +\frac{1}{2}\la 6S^z_iS^z_i-4\ra &\la
-S^-_i(2S^z_i-1)\ra & 0 & 0 & 0 & 0 & 0 \\
 -\la(2S^z_i-1)S^+_i\ra & 0 & 2\la S^+_iS^+_i\ra & 0 & 0 & 0 & 0 & 0 \\
0 & \la S^-_i(2S^z_i-1)\ra & -2\la S^-_iS^-_i\ra & 0 & 0 & 0 & 0 & 0 \\
3\la S^-_i(2S^z_i-1)\ra & -3\la(2S^z_i-1)S^+_i\ra & 0 & 0 & 0 & 0 & 0& 0
\end{array} \right) \;,}
\label{125}
\end{equation}

We now demonstrate that, if there is an eigenvector ${\bf L}^0$
with eigenvalue zero for the sub-matrix $\bf \Gamma_{ii}$, then
there is also a left eigenvector
of ${\bf \Gamma}$ corresponding to eigenvalue zero with the structure
\begin{equation}
{\bf L}^0=(0,...,0,{\bf L}^0_i,0,...,0) \ ,
\label{126}
\end{equation}
where, for spin $S=1$,
\begin{equation}
{\bf
L}_i^0=(L_{i,+}^0,L_{i,-}^0,L_{i,z}^0,L_{i,z+}^0,
L_{i,-z}^0,L_{i,++}^0,L_{i,--}^0,L_{i,zz}^0)
\label{127}
\end{equation}
Multiplying $\bf \Gamma$ from the left by ${\bf L}^0$ results in products of
${\bf L}_i^0$ with sub-matrices ${\bf \Gamma}_{ij}$.  The product with
${\bf \Gamma}_{ii}$ must be zero, since the diagonal blocks of $\bf \Gamma$
have the same structure as the monolayer matrix, Eq.~(\ref{11}).
For the off-diagonal blocks, ${\bf \Gamma}_{ij}$,
the product is also zero because of the regularity conditions
for layer $i$, derived from the fact that the
commutator Green's functions have to be regular at the origin;
see Refs.~\cite{EFJK99,FJK00} :
\begin{eqnarray}
\sum_{\alpha}\,L_{i\alpha}^0\;A_i^{\alpha,+}=0,\nonumber\\
\sum_{\alpha}\,L_{i\alpha}^0\;A_i^{\alpha,-}=0,\\
\sum_{\alpha}\,L_{i\alpha}^0\;A_i^{\alpha,z}=0.\nonumber
\label{128} \end{eqnarray}
Multiplying the non-diagonal matrix~(\ref{125}) from the left by the
eigenvector~(\ref{127}) and then applying the regularity
conditions Eqs.~(\ref{128}) yields zero.
Hence, we have shown that there are
as many zero eigenvalues of $\bf \Gamma$ as there are zero eigenvalues of
all of the diagaonal blocks ${\bf \Gamma}_{ii}$.  Since each diagonal
block has 2 zero eigenvalues (because each block has the same structure
as the monolayer matrix), there must be at least 2N zero eigenvalues
of the matrix $\bf \Gamma$.

Therefore,
apart from dimension, the equations determining the correlation
functions for the multi-layer system have the same form as
for the monolayer case:
\begin{equation}
{\bf C}={\bf R\;{\bf \cal E}\; L\;A}+{\bf R}^0\;{\bf L}^0\;{\bf C}\;.
\label{129}
\end{equation}
The matrices ${\bf R}$ and ${\bf L}$ have to be constructed from
the right and left eigenvectors corresponding to non-zero eigenvalues
as before, whereas
the matrices ${\bf R}^0$ and ${\bf L}^0$ are constructed from the
eigenvectors with eigenvalues zero.

\subsection*{4 Numerical results}
The results of the numerical calculations presented in this paper are meant
to demonstrate that our formulation in handling the single-ion anisotropy works
in practice. To this end we take the magnetic field components and the dipole
coupling constant to be zero and investigate the magnetization as a function of
the anisotropy strength and the temperature for a square monolayer with spin
$S=1$. In this case there is only a magnetization  $\la S^z \ra$ in z-direction.

\begin{figure}[htp]
\begin{center}
\protect
\includegraphics*[bb = 69 79 531 678,
angle=-90,clip=true,width=9cm]{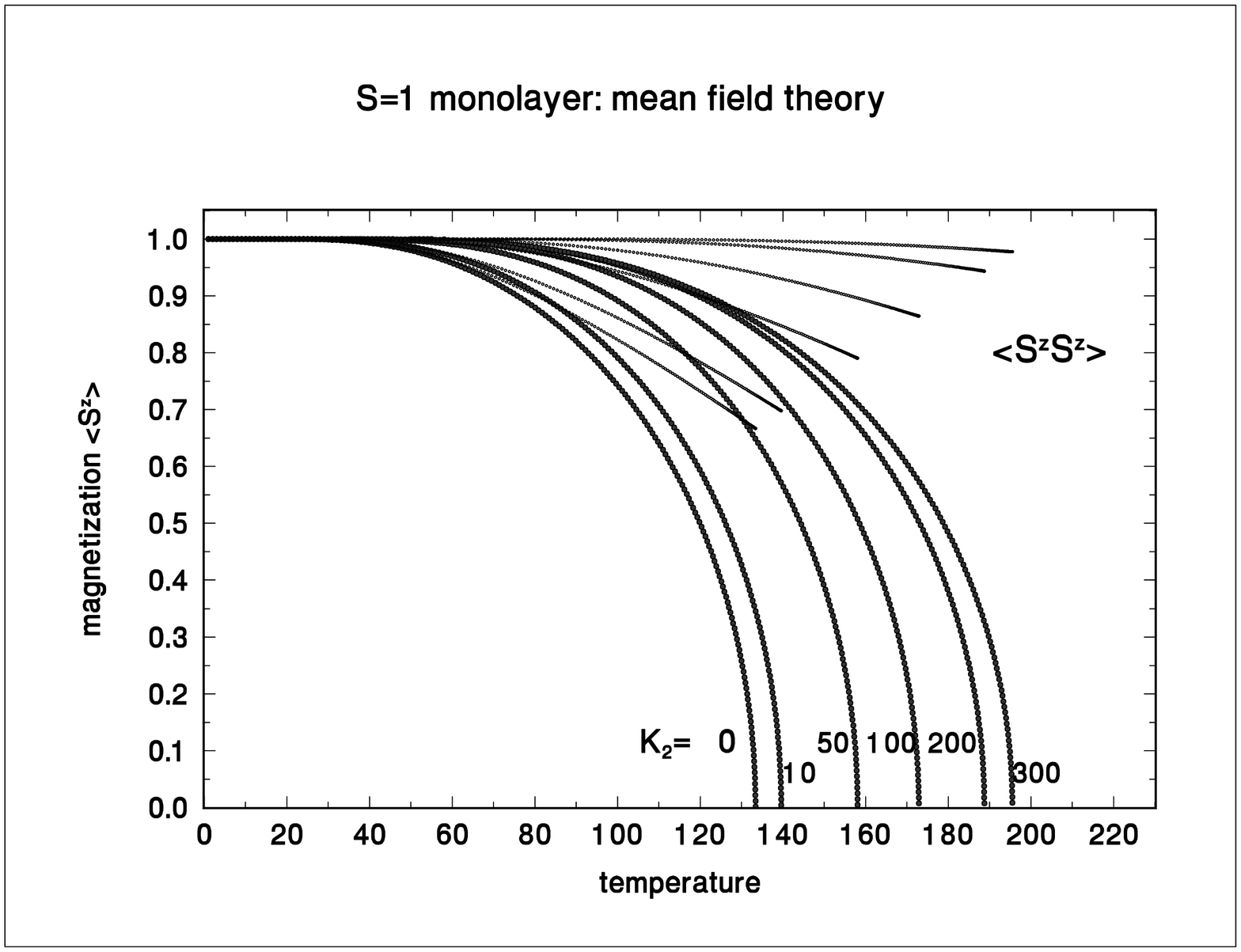}
\protect
\caption{\small{
Results of mean field calculations using either the mean field
limit
($\gamma_{\bf k}=0$) of the Green's function program or an exact diagonalization
of the corresponding mean field Hamiltonian. Both results are identical.
For a monolayer with $S=1$, the magnetization
component $\langle S^z\rangle$ and the correlation $\langle S^zS^z\rangle$
in MFT are shown as
functions of the temperature for anisotropy coefficients in the range
$ 0 < K_2 < 300 $; the exchange coupling strength is $J=100$.}}
\label{fig1}
\end{center}
\end{figure}

In Fig. 1 we show results of mean field (MFT) calculations for $\la S^z\ra$ and
$\la S^zS^z\ra$ as a function of the temperature for different anisotropies in
the range of $0<K_2<300$ obtained in two ways. The first is an
exact
diagonalization of the mean field Hamiltonian, which is possible because of its
one-body nature. If our Green's function theory (GFT) for the anisotropy term
is exact, calculations with the Green's function program in the mean field
limit (no momentum dependence on the lattice: $\gamma_{\bf k}=0$ of Eq.
(\ref{12})) should give identical results. This is indeed the case; both
results are indistinguishable in Fig. 1. The precise agreement of these very
different methods of calculations provides a check on the numerical
procedures.

\begin{figure}[htp]
\begin{center}
\protect
\includegraphics*[bb = 69 79 531 678,
angle=-90,clip=true,width=9cm]{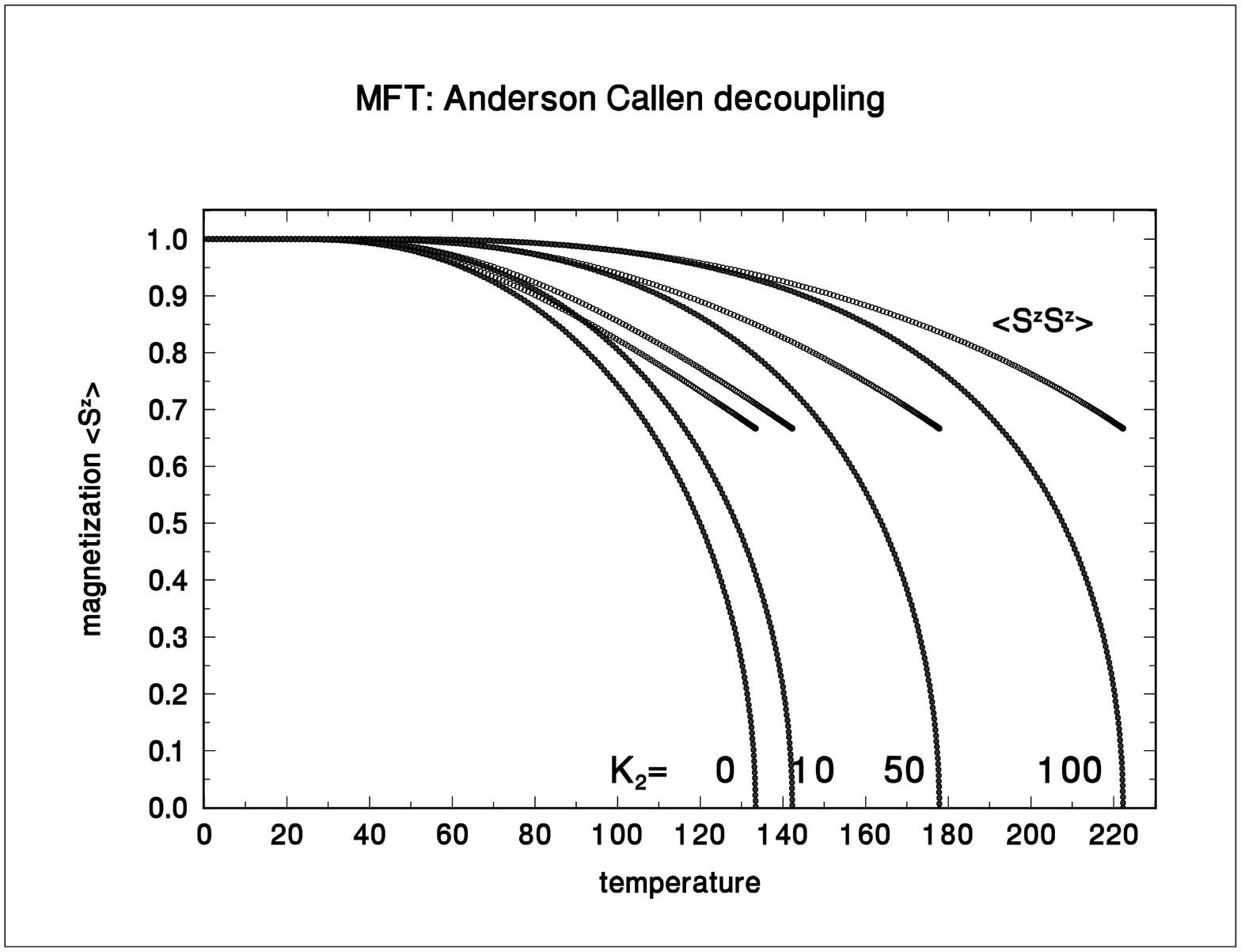}
\protect
\caption{ \small{
The figure displays results of the MFT limit of a GFT with the Anderson-Callen
decoupling, demonstrating the difference from the exact results of Fig. 1. The
magnetization $\la S^z\ra$ and correlation $\la S^zS^z\ra$
are shown only up to $K_2=100$,
where already the differences are large; results for $K_2>100$ lie outside the
temperature scale of the figure. Note that the values for $\la S^zS^z\ra=2/3$
contrast with the exact results.}}
\end{center}
\end{figure}
Fig.~2 presents the results of a MFT calculation using the Anderson-Callen
decoupling of the single-ion anisotropy terms.
The shortcoming of this decoupling is seen by comparing
with the exact results of Fig.~1. One observes that, up to $K_2=10$,
the approximate calculation overshoots the exact one only slightly, but
with increasing $K_2$ the disagreement becomes worse and worse.
The results for $K_2>100$ lie outside the frame of the figure.

In the MFT results of Figs.~1 and~2 the well-known shortcoming of MFT is
evident, the {\em violation} of the Mermin-Wagner theorem: there is a finite
Curie temperature for vanishing anisotropy:
$T_{Curie}^{MFT}(K_2=0)=\frac{4}{3}J=133.33$ for an exchange
coupling strength of $J=100$. For arbitrarily
large values of the anisotropy, the Curie temperature in MFT is obtained
analytically: $T_{Curie}^{MFT}(K_2\rightarrow \infty) =S^2qJ/(S(S+1))=200$ for
$S=1, J=100$ and $q=4$ (q is the coordination number of a square lattice).
This limit is almost reached numerically for $K_2=300$ as can be seen in
Fig.~1.

Our Green's function theory with the RPA-like treatment of the exchange terms
fulfills the Mermin-Wagner theorem:
$T_{Curie}\rightarrow 0$ for
$K_2\rightarrow 0$.

\begin{figure}[htp]
\begin{center}
\protect
\includegraphics*[bb = 69 79 531 678,
angle=-90,clip=true,width=9cm]{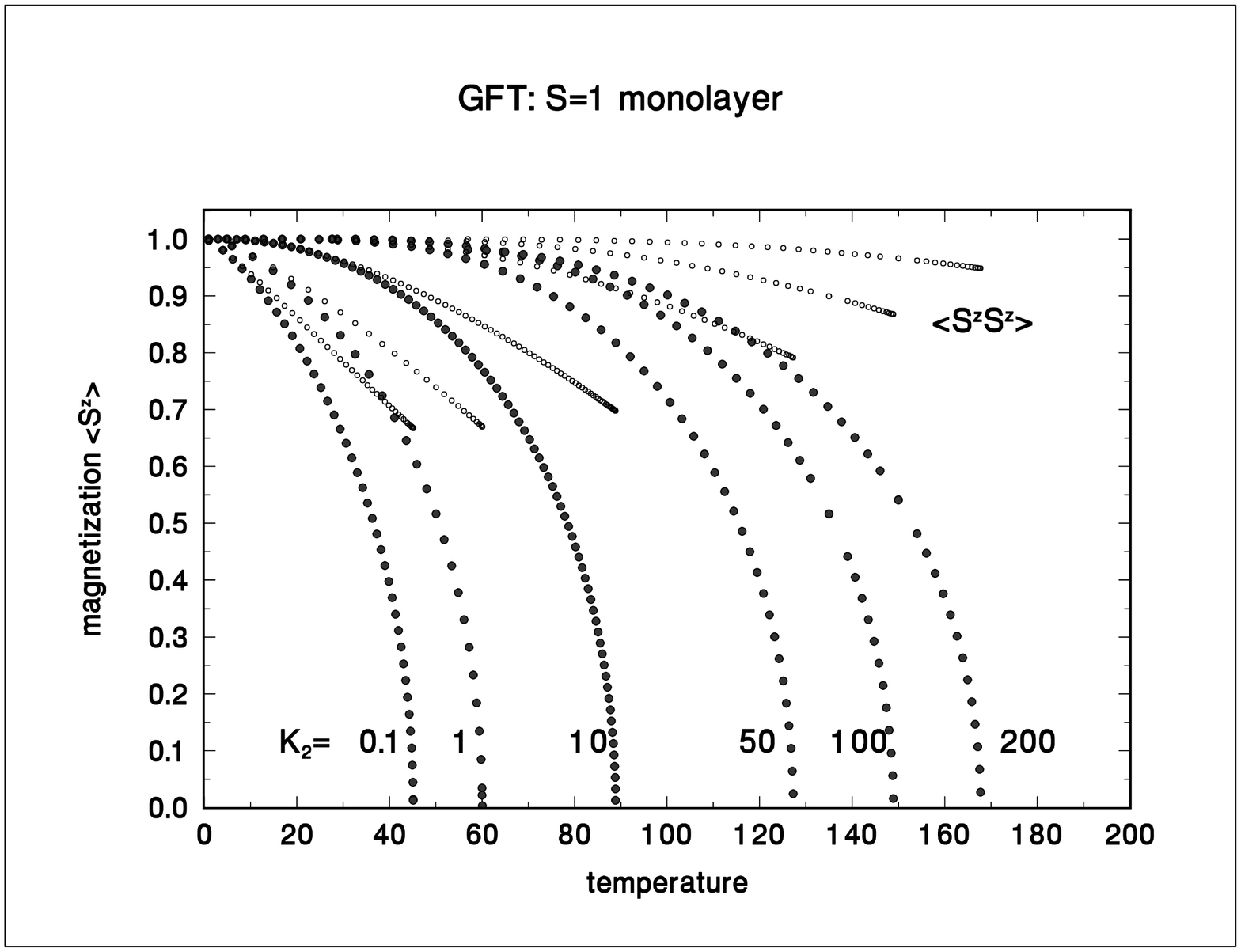}
\protect
\caption{\small{
 Results of the Green's function theory (with RPA-like
decouplings of the exchange terms) with the same input as in Fig.1 are
shown for $\langle S^z\rangle$ and $\langle S^zS^z\rangle$ as functions of the
temperature for various anisotropy coefficients $K_2$. Note the significant
differences from the mean field results of Fig.1.}}
\end{center}
\end{figure}
Comparison of the MFT results of Fig.~1 with the GFT results
in Fig.~3 reveals major differences between MFT and GFT with respect to
the temperature dependence of $\langle S^z\rangle$ for different
anisotropies $K_2$, particularly in the low temperature region and for small
anisotropies.
For large anisotropies it can be shown analytically that the full Green's
function theory approaches the MFT limit, $T_{Curie}=200$, when the anisotropy
becomes arbitrarily large (see Appendix B). This
is physically reasonable because, in the large anisotropy limit, GFT
approaches the Ising
limit, and, for the Ising model, a RPA treatment is identical with the mean
field
approach. The results of the exact treatment of the single-ion
anisotropy term shown in Fig.~3  represent a significant improvement
over the decoupling of this term proposed by Anderson and Callen \cite{AC64}
and the different decoupling of Lines\cite{Lin67},
both of which yield a diverging Curie temperature $T_{Curie}\rightarrow
\infty$
for $K_2\rightarrow \infty$. (See also Appendix B of
Ref. \cite{FJK00} in this connection.)

\begin{figure}[htp]
\begin{center}
\protect
\includegraphics*[bb = 69 79 531 678,
angle=-90,clip=true,width=9cm]{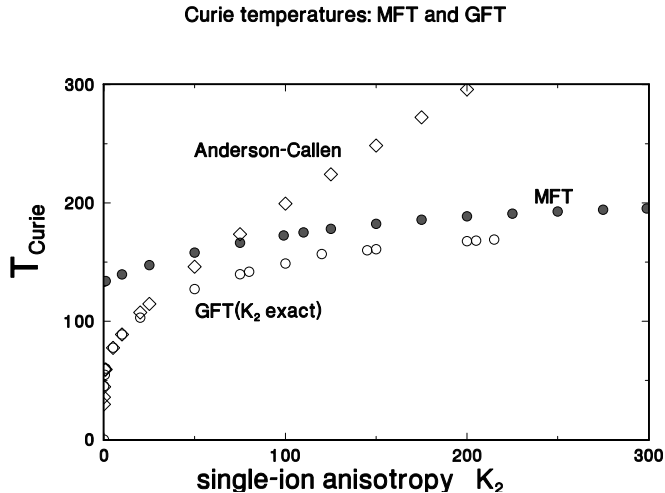}
\protect
\caption{\small{
Comparison of the Curie
temperatures calculated with the present exact treatment of the anisotropy
terms, the Anderson-Callen decoupling\cite{FJKE01}
and MFT. The first two approaches fulfill the Mermin-Wagner
theorem:
$T_C\rightarrow 0$ for $K_2\rightarrow 0$, whereas the MFT result does not.
For large anisotropies ($K_2\rightarrow\infty$), the new model
approaches slowly the mean
field results, as it should do, whereas the Anderson-Callen decoupling
procedure leads to a diverging $T_{C}$}}
\end{center}
\end{figure}
To show the difference between the new model and the Anderson-Callen
decoupling more clearly, we compare in Fig. 4 the Curie temperatures
obtained from MFT,
the new Green's function theory, and the Green's
function theory with the Anderson-Callen decoupling of
Refs.~\cite{FJK00,FJKE01}. For small anisotropies,  there is only
a slight difference between the two GFT results which, in contrast to MFT,
obey the Mermin-Wagner theorem. However, on increasing
the anisotropies, the GFT results deviate from one another significantly:
for $K_2\rightarrow \infty$, the Anderson-Callen
result diverges, whereas the exact treatment approaches the MFT limit, albeit
very slowly.

The difference between the exact treatment of the anisotropy terms and the
approximate Anderson-Callen decoupling is further demonstrated in Fig.~5, where
the magnetizations $\la S^z\ra$ as a function of the temperature for different
values of $K_2$ are compared. We see that, for small anisotropies, there is
rather good agreement, which, however, worsens as
$K_2$ increases.
Another difference concerns the second moments $\la S^zS^z\ra$, which, in the
case
of the Anderson-Callen decoupling, approach the value
$\la S^zS^z\ra(T\rightarrow T_{Curie})=2/3$ (see Ref.~\cite{FJK00}), whereas
in the exact treatment, the values of
$\la S^zS^z\ra(T\rightarrow T_{Curie})$ are
larger than $2/3$. This is as it should be, because, as shown in Appendix~B,
$\la S^zS^z\ra\rightarrow 1$ for $K_2\rightarrow\infty$.

\begin{figure}[htp]
\begin{center}
\protect
\includegraphics*[bb = 69 79 531 678,
angle=-90,clip=true,width=9cm]{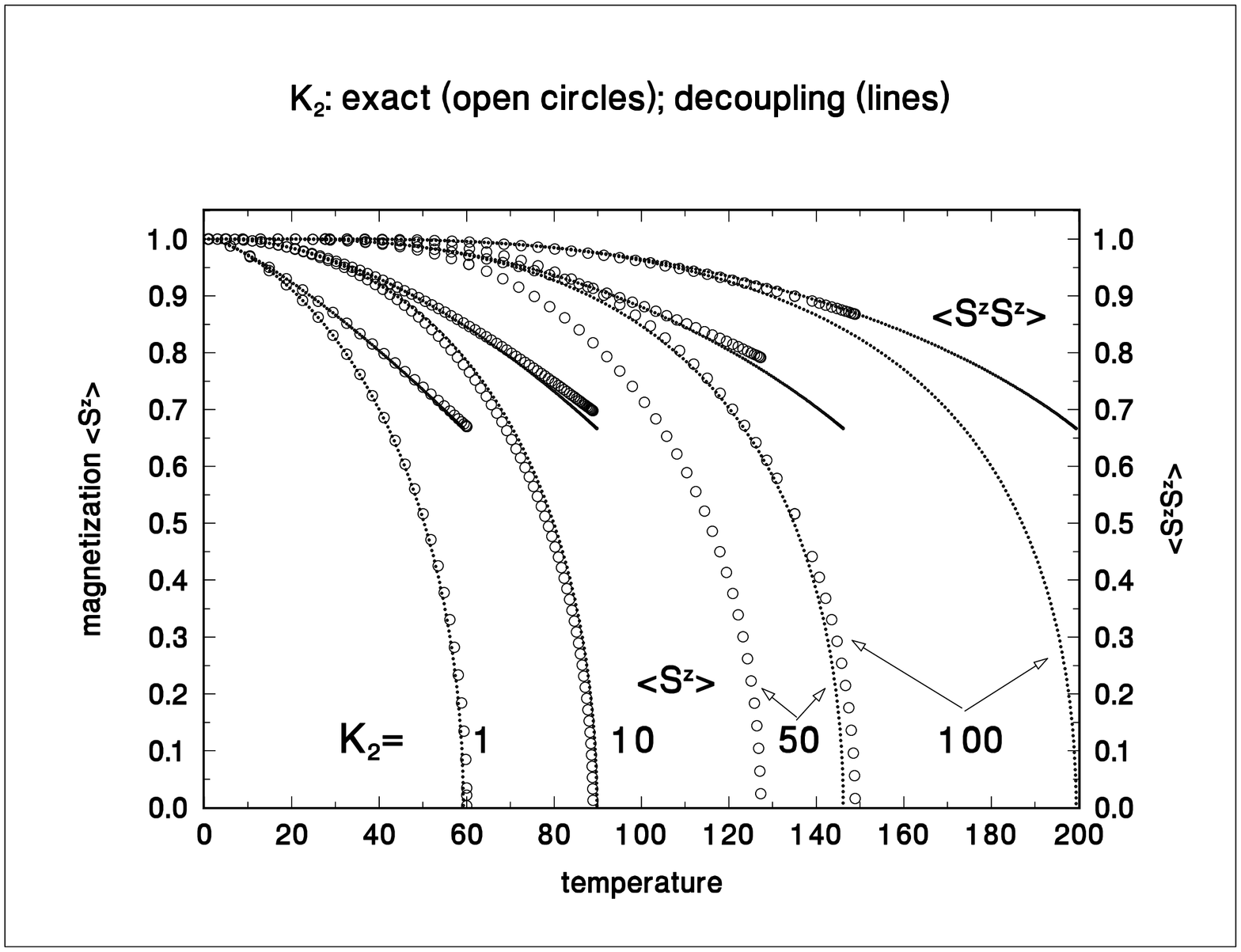}
\protect
\caption{\small{
Comparison of GFT calculations for $\la S^z\ra$ and $\la S^zS^z\ra$ as
functions of the temperature using the exact treatment of the anisotropy (open
circles) and the Anderson-Callen decoupling used in in Refs.~\cite{FJK00,FJKE01}
(small dots).}}
\end{center}
\end{figure}
\subsection*{5 Conclusions}

We have presented a formal theory for the magnetization of thin
ferromagnetic films on the basis of many-body Green's function theory
within a Heisenberg model with anisotropies. The essential improvement over
our previous work\cite{FJK00,FJKE01} is the exact
treatment of the single-ion anisotropy brought about by the introduction of
higher-order Green's functions. Previously, the anisotropy term was treated by
approximate decoupling procedures only at the level of the lowest-order Green's
functions. The exchange interaction terms are decoupled using an RPA-like
approach. We did not try to go beyond RPA
since our comparison with `exact' quantum Monte Carlo results has shown this to
be a very good approximation\cite{EFJK99}.

Numerical calculations of the magnetization as a function of the
temperature for various anisotropies $K_2$
(no external field, no dipole-coupling) demonstrate the
superiority of the new spin wave model over MFT. In particular,
there is no violation of the Mermin-Wagner theorem. The Anderson-Callen
decoupling used in our previous work gives results close to those of the
new model when the anisotropy $K_2$ is small but, as the anisotropy
increases, the difference between the two approaches becomes larger:
the new model approaches the MFT limit as it should do, whereas the
Curie temperature from the Anderson-Callen decoupling diverges.

Our new formulation should allow a future investigation of the reorientation
problem when switching on the magnetic field and/or the dipole coupling.
The treatment of multi-layer systems and spin $S>1$ should be possible.

We are indebted to A. Ecker and P.J. Jensen for discussions.

\subsection*{Appendix A: The curve-following procedure}
Consider a set of $n$ coupled equations characterised by $m$
parameters $\{P_i;i=1,2\ldots,m\}$
and $n$ variables $\{V_i;i=1,2,\ldots,n\}$:
\begin{equation}
   S_i({\bf P}[m];{\bf V}[n]) = 0, {i=1,\ldots,n}/ .
\end{equation}
In our case, the parameters are
the temperature, the magnetic field components, the dipole coupling strengths,
the anisotropy strengths, etc; the variables are the spin-correlations.
The coupled equations are obtained by defining the $S_i$ from the
$n$ self-consistency equations for the correlations vector
$\bf v$ (Eq.~(\ref{106})):
\begin{eqnarray*}
  {\bf S} & = & {\bf v} - \bf u_c^{-1}(R{\cal E}LA + R^0L^0(u_c v - u_c^0) - u_c^0).
\end{eqnarray*}

For fixed parameters ${\bf P}$, we look for solutions $S_i=0$
at localised points, ${\bf V}[n]$, in the n-dimensional space. If now one of
the parameters $P_k$ is considered to be an additional
variable $V_\circ$ (in this paper, the
temperature is taken as the variable),
then the solutions to the coupled equations define
curves in the $(n+1)$-dimensional space ${\bf V}[n+1]$. From here on, we
denote the points in this space by
$\{V_i;i=0,1,2,...\ldots,n\}$. The curve-following
method is a procedure for generating these solution-curves point by
point from a few closely-spaced points already on a curve; i.e. the method
generates a new solution-point from
the {\em approximate} direction of the curve in the vicinity of
a new {\em approximate} point. This is done by an iterative procedure
described below. If no points on the curve are known, then
an approximate solution point and an approximate direction must be
estimated before applying the iterative procedure to obtain the first
point on the curve. A second point can then be obtained in the same
fashion. If at least two solution-points are available, then the new
approximate point can be extrapolated from them and the approximate
direction can be taken as the tangent to the curve at the last point.

The iterative procedure for finding a better point,
${\bf V}$, from an approximate point, ${\bf V}^{\circ}$, is now described.
One searches for the isolated solution-point
in the $n$-dimensional subspace perpendicular to the approximate direction,
which we characterise by a unit
vector, $\widehat{\bf u}$.
The functions $S_i$ are expanded up to first order in the corrections
about the approximate point, ${\bf V^{\circ}}$:
\begin{equation}
   S_i({\bf V}) = S_i({\bf V^{\circ}}) + \sum_{j=0}^n \dfdx{S_i}{V_j}^{\circ}
  \Delta V_j ,
\end{equation}
where $\Delta V_j=V_j-V_j^{\circ}$.
At the solution, the $S_i$ are all zero, whereas at the approximate
point ${\bf V^\circ}$ the functions have non-zero values, $S_i^\circ$; hence,
one must solve for the corrections $\Delta V_j$ for which the left-hand side
in the above equation is zero:
 \begin{equation}
   \sum_{j=0}^n \dfdx{S_i}{V_j}^{\circ} \Delta V_j = -S_i^\circ;
\{i=1,2,\ldots,n\}.
 \end{equation}
These $n$ equations are supplemented by the constraint requiring the
correction to be perpendicular to the unit direction vector:
\begin{equation}
  \sum_{j=0}^n\widehat{u_j} \Delta V_j = 0.
\end{equation}
This improvement algorithm in the subspace is repeated until each of
the $S_i^\circ$ is sufficiently small. In practice we required that
$\sum_i{(S_i^\circ)^2 \leq \epsilon}$, where we took $\epsilon=10^{-16}$. If there is no convergence,
the extrapolation
step-size used to obtain the original ${\bf V}^{\circ}$ is halved,
a new extrapolated point obtained, and the improvement algorithm repeated.

The curve-following method is quite general and can be applied to any
coupled equations characterised by differentiable functions. By utilizing
the information about the solution at neighbouring points, the method is
able to find new solutions very efficiently, routinely converging after
a few iterations once two starting points have been found.

\subsection*{Appendix B: Curie temperature for $K_2\rightarrow\infty$}
We show analytically that the Curie temperature of the Green's function theory
with the exact treatment of the anisotropy for a square-lattice monolayer
with S=1 approaches the
mean field value when the anisotropy coefficient goes to infinity, whereas the
Anderson-Callen decoupling leads to a divergence in this limit.

For the case of a single magnetic direction,
the $8\times 8$ problem of Eq.~(\ref{10}) reduces
to a $2\times 2$ problem for the Green's functions
$G_{ij}^{+,-}=\la\la S_i^+;S_j^-\ra\ra$ and
$G_{ij}^{z+,-}=\la\la (2S_i^z-1)S_j^+;S_j^-\ra\ra$.
For this special case, it is possible to derive analytical
expressions for the correlations $\la S^z\ra$ and $\la S^zS^z\ra$:
\begin{eqnarray}
\la S^-S^+\ra&=&2-\la S^z\ra-\la S^zS^z\ra =\frac{1}{\pi^2}\int_0^\pi
dk_x\int_0^\pi dk_y \Big\{ \frac{1}{\omega^+-\omega^-}\nonumber\\
& &\Big[
(2\la S^z\ra(\omega^+-\la S^z\ra J_0)+K_2(6\la
S^zS^z\ra-4))\frac{1}{e^{\beta\omega^+}-1}\nonumber\\
& &-(2\la S^z\ra(\omega^--\la S^z\ra J_0)+K_2(6\la
S^zS^z\ra-4))\frac{1}{e^{\beta\omega^-}-1}
\Big]\Big\}
\label{b1}
\end{eqnarray}
\begin{eqnarray}
\la S^-(2S^z&-&1)S^+\ra=\la S^z\ra-\frac{1}{2}(6\la S^zS^z\ra-4)
=\frac{1}{\pi^2}\int_0^\pi
dk_x\int_0^\pi dk_y \Big\{ \frac{1}{\omega^+-\omega^-}\nonumber\\
& &\Big[
((6\la S^zS^z\ra-4)(\omega^+-\la S^z\ra J_0)+K_2(2\la
S^z\ra))\frac{1}{e^{\beta\omega^+}-1}\nonumber\\
& &-((6\la S^zS^z\ra-4)(\omega^--\la S^z\ra J_0)+K_2(2\la
S^z\ra))\frac{1}{e^{\beta\omega^-}-1}
\Big]\Big\}
\label{b2}
\end{eqnarray}
with
\begin{equation}
\omega^\pm=\frac{1}{2}\la S^z\ra(J_0-J_{\bf k})\pm\sqrt{K_2^2-\frac{1}{2}(6\la
S^zS^z\ra-4)K_2J_{\bf k}-\frac{1}{4}\la S^z\ra^2 }.
\label{b3}
\end{equation}
At the Curie temperature, $\la S^z\ra\rightarrow 0$, so that the equation
for $\omega^\pm$ becomes
\begin{equation}
\omega^\pm(\la S^z\ra=0)=\pm \omega^0=\pm\sqrt{K_2(K_2-\frac{1}{2}(6\la
S^zS^z\ra)-4)J_{\bf k})},
\label{b4}
\end{equation}
Equation~(\ref{b1}) then reduces to
\begin{equation}
2-\la S^zS^z\ra=(6\la S^zS^z\ra-4)\frac{1}{\pi^2}\int_0^\pi
dk_x\int_0^\pi dk_y \frac{K_2}{2\omega^0}\coth(\beta\omega^0/2).
\label{b5}
\end{equation}
For large $K_2$,
$\omega^0=K_2\sqrt{1-(6\la S^zS^z\ra-4)J_{\bf k}/(2K_2)}\simeq K_2$.
Passing to
the limit $K_2\rightarrow \infty$, one obtains from
Eq.~(\ref{b5}) at $T_{Curie}$
\begin{equation}
\la S^zS^z\ra=1.
\label{b6}
\end{equation}
Now, expanding Eq.~(\ref{b2}) around $\la S^z\ra=0$, and comparing the
coefficients of $\la S^z\ra$ of the resulting equation, one has at $T_{Curie}$
\begin{eqnarray}
1=\frac{1}{\pi^2}\int_0^\pi
dk_x\int_0^\pi dk_y& &\Big\{(\frac{K_2}{\omega^0}-\frac{J_{\bf k}(6\la
S^zS^z\ra-4)}{4\omega^0}) \coth(\frac{\beta\omega^0}{2})\nonumber\\
&+&(\frac{1}{2}J_{\bf k}-J_0)\beta(6\la
S^zS^z\ra-4)\frac{e^{\beta\omega^0}}{(e^{\beta\omega^0}-1)^2}\Big\}.
\label{b7}
\end{eqnarray}
Noting that $\int_0^\pi dk_x\int_0^\pi dk_y J_{\bf k}=0$ and that
$\omega^0\simeq K_2$ for large $K_2$, one obtains from Eq.~(\ref{b7})
\begin{equation}
1-\coth(\frac{\beta K_2}{2})=-J_0\beta (6\la S^zS^z\ra-4))
\frac{e^{\beta K_2}}{(e^{\beta K_2}-1)^2}.
\end{equation}
Again, goint to the limit $K_2\rightarrow\infty$
and using Eq.~(\ref{b6}),
one obtains for the Curie temperature
\begin{equation}
T_{Curie}=J_0\frac{1}{2}(6\la S^zS^z\ra-4)=J_0=4J
\end{equation}
This is just the MFT result! This is physically reasonable because a large
anisotropy approaches the Ising limit, and the RPA for the Ising model is
identical to its mean field treatment.

This is in contrast to the result of the decoupling procedure.
In Appendix B of Ref.~\cite{FJK00} we have shown that the Anderson-Callen
decoupling of the anisotropy term leads for a square monolayer to a Curie
temperature
\begin{equation}
T_{Curie}\simeq \frac{8\pi J/3}{\ln(1+3\pi^2J/K_2)},
\end{equation}
which diverges for $K_2\rightarrow\infty$!

\end{document}